\documentclass[a4paper,fleqn,usenatbib,useAMS]{mnras}

\usepackage{color}
\usepackage{graphicx}
\usepackage{amsmath}	
\usepackage{amssymb}	
\usepackage{multicol}        
\usepackage{bm}		
\usepackage{pdflscape}	
\usepackage{subcaption}
\usepackage{comment}

\usepackage[T1]{fontenc}
\usepackage{ae,aecompl}
\usepackage{enumitem}
\usepackage{colortbl}

\usepackage{newtxtext,newtxmath}

\title[Fornax globular cluster distributions: implications for the cusp-core problem]{Fornax globular cluster distributions: implications for the cusp-core problem}

\author[P. Boldrini, R. Mohayaee, J. Silk]{Pierre Boldrini$^{1}$\thanks{Contact e-mail: \href{mailto:boldrini@iap.fr}{boldrini@iap.fr}}, {Roya Mohayaee$^{1}$}, {Joseph Silk$^{1,2,3}$}
\\
$^{1}$Sorbonne Universit\'e, CNRS, UMR 7095, Institut d'Astrophysique de Paris, 98 bis bd Arago, 75014 Paris, France\\
$^{2}$Department of Physics and Astronomy, The Johns Hopkins University, Baltimore MD 21218, USA\\
$^{3}$Beecroft Institute for Particle Astrophysics and Cosmology, Department of Physics, University of Oxford, Oxford OX1 3RH, UK}
\date{Accepted 2019 February 24. Received 2019 February 18; in original form 2018 July 2.}

\pubyear{2017}

\begin{document}
\label{firstpage}
\pagerange{\pageref{firstpage}--\pageref{lastpage}}
\maketitle

\begin{abstract}
We re-investigate the Fornax cusp-core problem using observational results on the spatial and mass distributions of globular clusters (GCs) in order to put constraints on the dark matter profile. We model Fornax using high resolution N-body simulations with entirely live systems, i.e.  self-gravitating systems composed of stars and dark matter, which account correctly for dynamical friction and tidal effects between Fornax and the globular clusters. We test two alternative hypotheses, which are a cored and a cuspy halo for Fornax by exploring a reasonable range of initial conditions on globular clusters. For Fornax cored dark matter halo, we derive a lower limit on the core size of $r_{c}\gtrsim$ 0.5 kpc. Contrary to many previous works, we show also that for different initial conditions, a cuspy halo is not ruled out in our simulations based on observations of Fornax globular clusters. 
\end{abstract}





\section{Introduction}

Cold dark matter (CDM) cosmology is the standard paradigm of structure formation in the Universe. However, there are a few issues on small scales in the context of this model such as the missing satellites problem, the too-big-to-fail problem, the cusp-core problem and the satellite disk problem (see e.g. \cite{2017ARA&A..55..343B} for a recent work). In this work, we focus on the cusp-core problem. Measurements of galaxy rotation curves and dynamical models of dwarf spheroidal galaxies have revealed that the density profile of the dark matter (DM) halo is constant at the centres of dwarf galaxies and therefore corresponds to a cored profile \citep{1994Natur.370..629M,1995ApJ...447L..25B,2001ApJ...552L..23D,2003ApJ...583..732S,2005AJ....129.2119S,2011ApJ...742...20W}. In contrast, N-body simulations have generally predicted a steep power-law mass-density distribution at the centre of CDM halos, that is a cuspy profile \citep{1997ApJ...490..493N,1997ApJ...477L...9F,1998ApJ...499L...5M}. 

Dwarf spheroidal (dSphs) galaxies are among the most dark matter-dominated galaxies in the Universe \citep{2013NewAR..57...52B,2013pss5.book.1039W}. In some dSphs, dark matter constitutes 90$\%$ or more of the total mass, even at the centre of the galaxy, so the dynamics is determined entirely by the gravitational field of the dark matter. Therefore, these systems  provide an excellent laboratory to study dark matter in the context of galaxy formation and evolution. Fornax is the most massive of the Milky Way dSphs and is the only one to have five globular clusters orbiting in a dense background of dark matter. For instance, another major dSph, such as the Sagittarius dSph, has many globular clusters. Four clusters are found in its main body (M 54, Arp 2, Terzan 7 and Terzan 8) \citep{1995AJ....109.2533D} and a few halo clusters have been associated with the stream across the sky (\cite{2015A&A...579A.104S} and references therein). The study of globular cluster dynamics can place powerful constraints on the Fornax dark matter halo type. 

One apparent paradox about these clusters is that we do not expect to see any of them because they should have sunk to the centre of Fornax due to dynamical friction \citep{1943ApJ....97..255C}. It is precisely because of this drag force that globular cluster are expected to sink to the centre of their host galaxy and form a nuclear star cluster \citep{1975ApJ...196..407T,1976ApJ...203..345T}. However, there is no bright stellar nucleus and we are still observing globular clusters orbiting in Fornax. This has become known as the Fornax timing problem \citep{2000ApJ...531..727O}. Since the dynamical friction force depends directly on the density of dark matter halo, the timing problem could be used to probe the cusp-core problem. Changing the density profile could delay dynamical friction. Simulations agree well with Chandrasekhar's analytic calculation for a cuspy halo such as Navarro-Frenk-White (NFW) halo, which is the most commonly used model for dark matter halos \citep{2006MNRAS.368.1073G}. Enhancement of  the infall time can be achieved by changing the density profile of the dark matter halo.

Numerous simulations have been performed to study the timing problem \citep{2000ApJ...531..727O,2006MNRAS.368.1073G,2006MNRAS.373.1451R,2012MNRAS.426..601C}. A live system is necessary to capture tidal stripping and dynamical friction. Thus, Fornax has to be modelled as a live galaxy with its globular clusters, i.e. a self-gravitating system composed of star and dark matter particles. Concerning the determination of the dark matter density distribution, previous simulations showed that the existence  of  five  globular  clusters  in  Fornax  is  inconsistent  with the hypothesis of a cuspy halo due to dynamical friction, because the globular clusters would have sunk into the centre of Fornax in a relatively short time \citep{2006MNRAS.370.1829S,2006MNRAS.368.1073G,2016MNRAS.461.4335A,2017MNRAS.464.3060A}. According to \cite{2006MNRAS.368.1073G}, a solution to the timing problem is that Fornax has a cored dark matter halo. In this model, globular clusters stop sinking at the core radius because of a resonance/scattering effect. Later work provided further support for a large dark matter core for Fornax \citep{2012MNRAS.426..601C}. The globular clusters did not fall to the centre because  of `dynamical buoyancy' created by the Fornax core. It was also proposed that globular clusters did not form within Fornax, but that they have been accreted by Fornax, which has a small cored halo. 

In this work, we re-investigate the Fornax cusp-core problem using N-body simulations with entirely live objects i.e. fully composed of particles, in order to take into account correctly dynamical friction and tidal effects between Fornax and its globular clusters. The paper is organized as follows. Section 2 provides a clear description of the Fornax system and the N-body modelling. In Section 3, we present details of our numerical simulations. In Section 4, we present our results of simulations and discuss the implications of our result on the dark matter halo profile of Fornax.

\section{Fornax-globular cluster system}

\begin{table*}
\label{tab:landscape}
\begin{tabular}{cccccccccc}
 \hline
    Object & $M^{a}$ & $r_{c}^b$&  $r_{t}^{d,e}$ & $D_{p}^b$ & $D_{MW}^{a}$ \\
     & [$10^{5}$ $M_{\sun}$] & [pc]& [pc] & [kpc] & [kpc]  \\
    \hline
    Fornax & 382 $\pm$ 12 & 668  $\pm$ $34^c$ & - & - & 147 $\pm$ 4\\
    GC1 & 0.42 $\pm$ 0.10 & 10.03 $\pm$ 0.29  & 59.06 $\pm$ 1.70 & 1.6 & 147.2 $\pm$ 4.1  \\
    GC2 & 1.54 $\pm$ 0.28 & 5.81 $\pm$ 0.19  & 108.19 $\pm$ 3.54 & 1.05 & 143.2 $\pm$ 3.3\\
    GC3 & 4.98 $\pm$ 0.84 & 1.60 $\pm$ 0.07 & 108.17 $\pm$ 4.73 & 0.43 & 141.9 $\pm$ 3.9   \\
    GC4 & 0.76 $\pm$ 0.15 & 1.75 $\pm$ 0.18 & 115.62 $\pm$ 11.89 & 0.24 & 140.6 $\pm$ 3.2\\
    GC5 & 1.86 $\pm$ 0.24 & 1.38 $\pm$ 0.11 & 25.69 $\pm$ 2.05 & 1.43 & 144.5 $\pm$ 3.3 \\
    \hline
\end{tabular}
\caption{Observed data for the Fornax system. $r_c$ is the half-light radius for Fornax dwarf galaxy and the King model core radius for the globular clusters with their tidal radius $r_t$. $D_{p}$ is the projected distance of the globular cluster (GC) from the centre of Fornax.  The last column is the line-of-sight distance from Milky Way. References: (a) \protect\cite{2016A&A...590A..35D}, (b) \protect\cite{2003MNRAS.340..175M}, (c) \protect\cite{2010MNRAS.408.2364S}, (d) \protect\cite{1985IAUS..113..541W},(e) \protect\cite{1996AJ....111.1596S}.}
\label{go}
\end{table*}

The dSph galaxy Fornax is one of the most dark matter-rich satellites of the Milky Way  with a mass of about $10^{8}$ M$_{\sun}$ at a distance of around 147 kpc \citep{2016A&A...590A..35D}. Fornax contains five globular clusters with masses of about $10^{5}$ M$_{\sun}$ and the average projected distance of these clusters is about 1 kpc. Various details are given in Table~\ref{go}. In this section, we present the models for Fornax and its globular clusters that provide the initial conditions for our simulations. 

\subsection{Fornax modelling}

\begin{table}
\centering
  \label{tab:landscape}
  \begin{tabular}{cccccccc}
    \hline
    Model & Density profile & $r_{s}$  & $M_h$ & $r_{t}$ & $m_p$ \\
     &  &  [kpc] & [M$_{\sun}$] & [kpc] & [M$_{\sun}$]\\
    \hline
    $B_1$ & Burkert & 0.25 & $0.318\times10^{9}$ & 2.81 & 89\\
    $B_2$ & Burkert & 0.5 & $0.88\times10^{9}$ & 3.95 & 230\\
    $B_3$ & Burkert & 0.75 & $1.1\times10^{9}$ & 4.52 & 285\\
    $B_4$ & Burkert & 1 & $1.28\times10^{9}$ & 4.98 & 329\\
    $N_1$ & NFW & 0.5 & $0.6\times10^{9}$ & 3.87 & 160\\
    $N_2$ & NFW & 1.0 & $1.2\times10^{9}$ & 4.87 & 310\\
    $N_3$ & NFW & 1.5 & $1.6\times10^{9}$ & 5.25 & 410\\
    $N_4$ & NFW & 2 & $2.0\times10^{9}$ & 5.77 & 510\\
    \hline
\end{tabular}
\caption{Parameters for halo models used in the simulations. $r_{s}$ and $M_h$ are the scale radius (core radius for Burkert profile) in density profiles and mass for the halo respectively. $r_t$ is the tidal radius for Fornax estimated from Eq~(\ref{rad}). We run cored halos $B_n$ modelled by a Burkert profile and cuspy halos $N_n$ modelled by a NFW profile. The halo and the stellar component are represented by $N=10^{6}$ particles. $m_p$ is the mass resolution for our N-body realization.}
\label{ta}
\end{table}

We construct Fornax as a live galaxy composed of stars and dark matter particles only, since dSph galaxies contain little or no gas today. 

The Fornax stellar component is modelled due to the presence of the dSph core ($r_{0}=0.668$ kpc) by a Plummer profile \citep{1911MNRAS..71..460P}:
\begin{equation}
\rho(r)=\frac{3r_{s}^{2}M_{0}}{4\pi}(r^{2}+r_{s}^{2})^{-5/2}, 
\label{plu}
\end{equation}
where $r_{s}$ and $M_{0}$ are the scale parameter and the mass, respectively. 

For the dark matter halo of Fornax, we employ two different density profiles: NFW and Burkert profiles in order to deal with the cusp-core problem. 

For cuspy halos, we assume a NFW form \citep{1996ApJ...462..563N}: 
\begin{equation}
\rho(r) = \rho_{0}\left(\frac{r}{r_{s}}\right)^{-1}\left(1+\frac{r}{r_{s}}\right)^{-2},
\end{equation}
with the central density $\rho_{0}$ and scale length $r_{s}$. 

For cored halos, we assume that the dark matter is distributed in a spherical halo with a Burkert density profile \citep{1995ApJ...447L..25B,2000ApJ...537L...9S}:
\begin{equation}
\rho(r) = \frac{\rho_{0}r_{0}^{3}}{(r+r_{0})(r^{2}+r_{0}^{2})},
\end{equation}
where $r_{0}$ and $\rho_{0}$ are the core radius and the central density, respectively.

\begin{figure}
\centering
\includegraphics[width=0.47\textwidth]{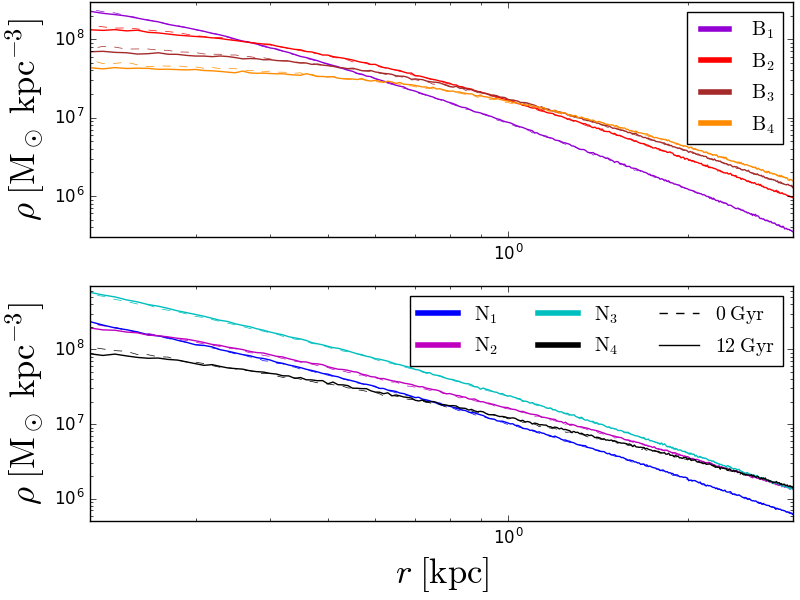}
\caption{Density distributions of cored ({\it top panel}) and cuspy ({\it bottom panel}) dark matter halo models. Radial density profiles at the beginning $T=0$ Gyr (solid line) and at the end of the simulation $T=12$ Gyr (dotted line), which are the same, show the stability of our halos. All halo parameters are summarized in Table~\ref{ta}.}
\label{dens2}
\end{figure}

In order to determine the halo parameters (see Table~\ref{ta}), we fitted the data of Fornax mass with a mass model that includes a stellar component and a dark halo: 
\begin{equation}
M_{\rm model}(r) = M_{*}(r) + M_{\rm h}(r),
\end{equation}
for all halos in Figure~\ref{dens1}. The data points correspond to the mass estimates by \cite{2011ApJ...742...20W} for two chemically distinct sub-populations. There are other mass estimators for Fornax \citep{2013MNRAS.429L..89A,2018MNRAS.481.5073E}. Plots of density profiles are illustrated in Figure~\ref{dens2}. In order to test halo stability during the simulation, we compared our halo profiles at the beginning ($T=0$ Gyr (solid line)) and at the end ($T=12$ Gyr (dotted line )) of the simulation, which match almost exactly for all radii.

\subsection{Globular cluster modelling}

\begin{figure}
\centering
\includegraphics[width=0.47\textwidth]{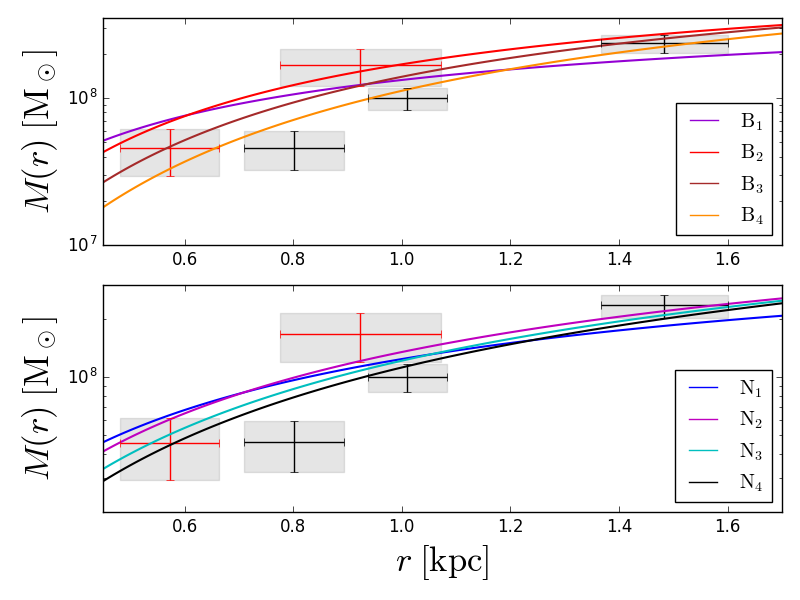}
\caption{Enclosed mass of cored ({\it top panel}) and cuspy ({\it bottom panel}) dark matter halo models. The data points correspond to the mass estimates by \protect\cite{2011ApJ...742...20W} for two chemically distinct subpopulations and by \protect\cite{2013MNRAS.429L..89A} for three distinct stellar subpopulations in the red giant branch. All halo parameters are summarized in Table~\ref{ta}.}
\label{dens1}
\end{figure}

\begin{figure}
\centering
\includegraphics[width=0.47\textwidth]{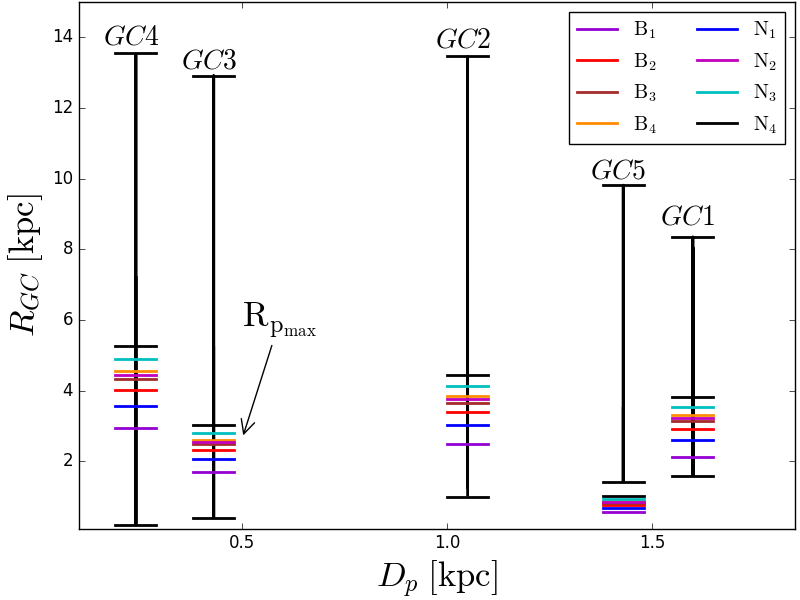}
\caption{All possible radial distances from the centre of Fornax for the five globular clusters, based on the observed line-of-sight distance with uncertainties, as a function of the projected distance. The minimal values correspond to the current projected distance $D_{p}$ of each globular cluster. The radial distance can be much bigger than the projected distance. The colour lines represent the maximum pericentre of observed globular clusters resulting from the fitted cluster tidal radius of Table~\ref{ta} for each halo models. We applied this last constraint only for high eccentric orbit with $e=0.9$.}
\label{radius}
\end{figure}

There are five surviving globular clusters orbiting in Fornax. Our N-body realizations of globular clusters assume a \cite{1962AJ.....67..471K} stellar density distribution,
\begin{equation}
\rho(r) = \rho_{0}\left[\left(1+\left(\frac{r}{r_k}\right)^2\right)^{-1/2} - C\right], \quad C=\left[1+\left(\frac{r_t}{r_k}\right)^2\right]^{-1/2},
\end{equation}
where $r_c$ and $r_t$ are the King and tidal radii, respectively. For most of the simulations, we chose a King radius $r_{k} = 1$ pc lower than the observed radius (see Table~\ref{dens1})
, because it is susceptible to increase through dynamical processes such as mass loss. The projected distances for the globular clusters are from 0.24 to 1.6 kpc, which is the minimum distance between globular clusters and the Fornax centre. Based on the observed line-of-sight distance with uncertainties (see Table~\ref{go}), as a function of the projected distance, we calculate all possible radial distances from the centre of Fornax for all globular clusters, which are summarized in Figure~\ref{radius}. According to this figure, the radial distance can be larger than the projected distance. We also estimate the maximum pericentre of observed globular clusters with eccentricity parameter $e=0.9$ thanks to the fitted cluster tidal radius of Table~\ref{dens1} for each halo models. These radii will be used as constraints on globular cluster final orbital radii (see Figure~\ref{radius}).

To generate the initial conditions, we use the numerical code, \textsc{nbodygen} \citep{2014MNRAS.442..160S}. This C++ code draws positions and velocities of each particle such that the resulting distribution follows the desired density profile $\rho(r)$. The code ensures that the final realization of the galaxy is in dynamical equilibrium.

\subsection{Milky Way tidal field}

According to Gaia DR2 data \citep{2018A&A...616A..12G}, the orbit of Fornax has an eccentricity of 0.29 and its pericentre is about 85.9 kpc. We estimate the tidal radius for Fornax using the equation of \cite{2006MNRAS.366..429R}:
\begin{equation}
r_{t}= r_p \left(\frac{M_{s}}{M_{g}(3+e)}\right)^{1/3},
\label{rad}
\end{equation}
where $r_p$, $M_s$, $M_g$ and $e$ are the percienter radius, the satellite mass, the galactic mass and the eccentricity respectively. We find tidal radii of 2.81-5.77 kpc, based on the range of masses for Fornax given in Table~\ref{ta} and using a total mass for the Milky Way of 2$\times10^{12}$ M$_{\sun}$. Within these radii, we can expect that tidal effects do not profoundly alter the structure and kinematics of Fornax. However, the Milky Way tidal field can have a significant impact on globular clusters on eccentric orbits (see Appendix B). Our host galaxy potential, based on the model of \cite{1991RMxAA..22..255A}, consists of a stellar bulge as a Plummer sphere \citep{1911MNRAS..71..460P}, a disc represented by the potential from \citep{1975PASJ...27..533M} and a spherical dark matter halo described by NFW profile \citep{1997ApJ...490..493N}. For this model, we used the revised parameters from \cite{2013A&A...549A.137I} (see their Table 1).

\section{N-body simulations}

Fornax is dominated by metal rich stars whereas globular clusters are dominated by metal-poor stars \citep{2016A&A...590A..35D}. If we compare the total mass of metal-poor ([Fe/H]<-2) clusters of (8.81 $\pm$ 0.92)$\times10^{5}M_{\sun}$ to the metal-poor stellar mass in Fornax of (44.9 $\pm$ 5.3)$\times10^{5}$ $M_{\sun}$, this yields you a mass fraction of 19.6 $\pm$ 3.1 $\%$, which implies that a large fraction of the metal-poor stars in Fornax still belong to the globular clusters \citep{2016A&A...590A..35D}. GC4 was excluded from this estimation, because this cluster is possibly more metal-rich than the other clusters. This high mass fraction of 19.6 $\pm$ 3.1 $\%$ suggests that each of these four surviving metal-poor globular clusters has likely lost several times $10^{5}$ $M_{\sun}$ due to dynamical processes such as dynamical friction, tidal effects and evaporation as a result of two-body relaxation. These processes act to destroy globular clusters on Gyr time-scales \citep{2001ApJ...561..751F,2007ApJS..171..101J}. Based on this hypothesis, we supposed that globular clusters were initially much more massive in the past and belonged to Fornax. Thus, the globular cluster initial radii have to be lower than the approximate Fornax tidal radii depending on the halo mass.

We ran 75 simulations with five globular clusters orbiting in all our halo models. The initial parameters are orbital radii $R_i$ = $[1.0, 1.5, 2.0, 2.5, 3.0]$ kpc and globular cluster masses $M_i$ = $[2.5, 5.0, 7.5, 10]$ $\times10^{5}M_{\sun}$. All runs were made with eccentricity parameters $e=0$ and $e=0.9$. They correspond to circular and high eccentric orbit, respectively, with an orbital velocity which depends on the density profile of Fornax $\rho(u)$. In order to put constraints on the dark matter halo, we consider only these two limit cases. In fact, the lifetime of clusters can be increased on high eccentric orbit, because tidal effects decrease with eccentricity (see Equation~(\ref{rad})).

We performed our simulations with the N-body code of \textsc{gadget2} (\cite{2005MNRAS.364.1105S}). We create, for each Fornax mass model, initial conditions for the Fornax-globular clusters system and evolve them  for 12 Gyr because the Fornax globular clusters are all dominated by ancient (>10 Gyr) populations of stars \citep{2016A&A...590A..35D}.

The halo and the stellar components are represented by $N=4\times10^{6}$ particles. The mass ratio for these two components determines the particle number for each component. Globular clusters are represented by about $10^{3}$ particles depending on the halo particle mass. In fact, we impose that the particle mass of all components is set to be equal in order to reduce numerical artefacts. Based on convergence tests of decaying radial distances and mass loss of the globular clusters (see Appendix A), the forces between all particles are softened with the same softening length of $\epsilon = 1$ pc. The softening length is similar to the King radius in order to maintain the dynamical stability of an isolated globular cluster. We did not use this radius as a constraint on halo models because we recognize that we do not have enough resolution in some halo models for the globular cluster, especially with a low initial masses.

In such systems, two mechanisms are responsible for orbital decay and mass loss: tidal effects and dynamical friction induced by the dark matter halo. The cluster mass plays an important role in its evolution and survival and dynamical friction is responsible for the orbital decay. These two processes compete with and regulate each other: orbital decay increases the tidal field, which reduces the globular cluster mass, and hence slows down the orbital decay. To describe cluster orbital decay, we calculated the distance between the cluster mass centre and Fornax mass centre at each snapshot, in order to get the orbital radius. In order to estimate the globular cluster mass loss, we count only bound particles. The dissolution times were defined to be the time when 95\% of the mass was lost from the globular cluster.

\section{Results}

We present and discuss our simulation results. To analyze our data and extract our results, we use a \textsc{python} module toolbox, \textsc{pnbody 4.0} \citep{2013ascl.soft02004R}.

\begin{figure}
\centering
\includegraphics[width=0.47\textwidth]{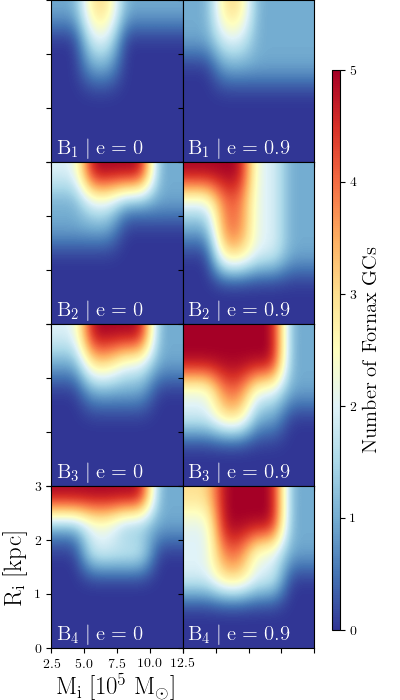}
\caption{Number of Fornax globular clusters reproduced by cored halo models $B_n$  with both eccentricity parameters $e=0$ and $e=0.9$ for our range of initial conditions. The initial parameters are orbital radii $R_i$ = $[1.0, 1.5, 2.0, 2.5, 3.0]$ kpc and globular cluster masses $M_i$ = $[2.5, 5.0, 7.5, 10]$ $\times10^{5}M_{\sun}$. $B_2$, $B_3$ and $B_4$ models can reproduce observations for a relevant range of initial cluster orbital radii and masses.}
\label{sB}
\end{figure}

\begin{figure}
\centering
\includegraphics[width=0.47\textwidth]{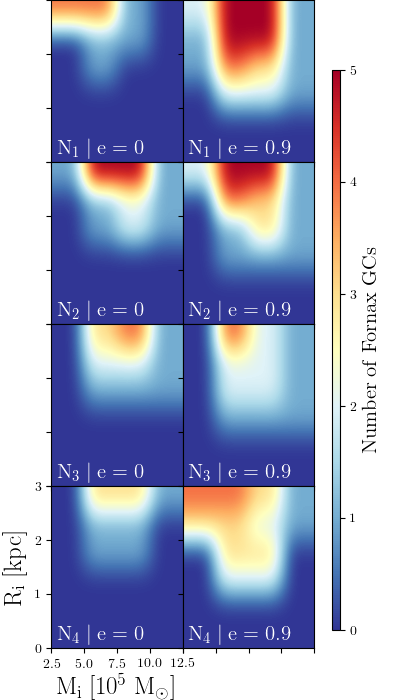}
\caption{Same as Figure~\ref{sB}, except for cuspy halo models $N_n$. $N_2$ ($r_{s}=$ 1 kpc) model can reproduce globular cluster distributions compatible with observations for both eccentric parameters.}
\label{sN}
\end{figure}

\begin{figure*}
\centering
\begin{minipage}[t]{8.4cm} 
\centering
\includegraphics[width=8cm]{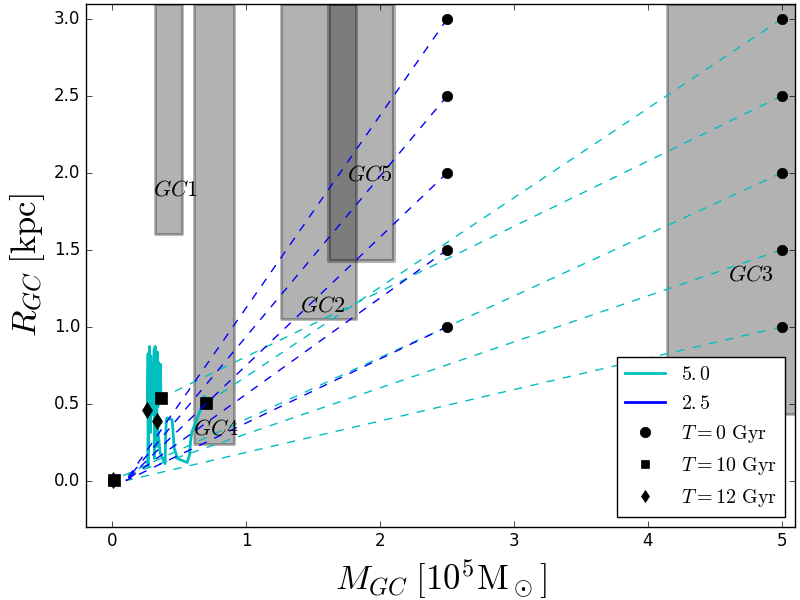}
\end{minipage} 
\begin{minipage}[t]{8.4cm}
\centering 
\includegraphics[width=8cm]{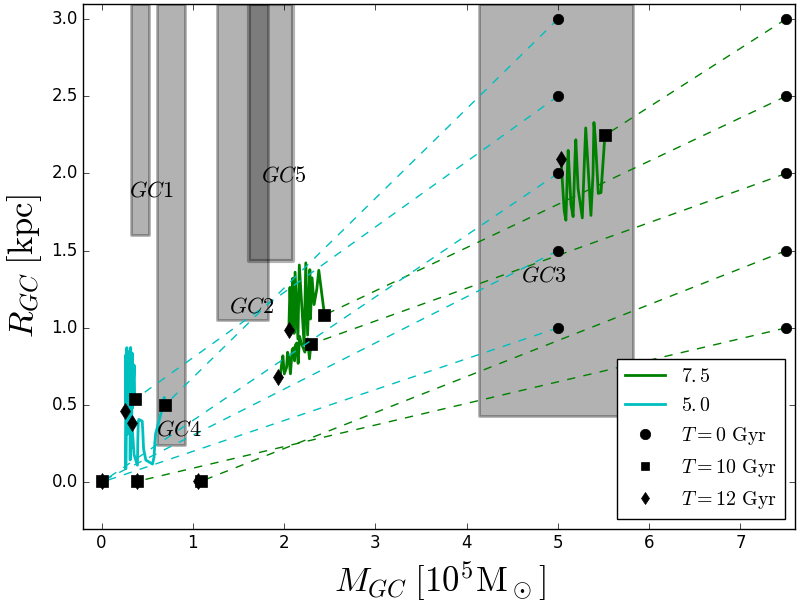}
\end{minipage}
\begin{minipage}[t]{8.4cm} 
\centering
\includegraphics[width=8cm]{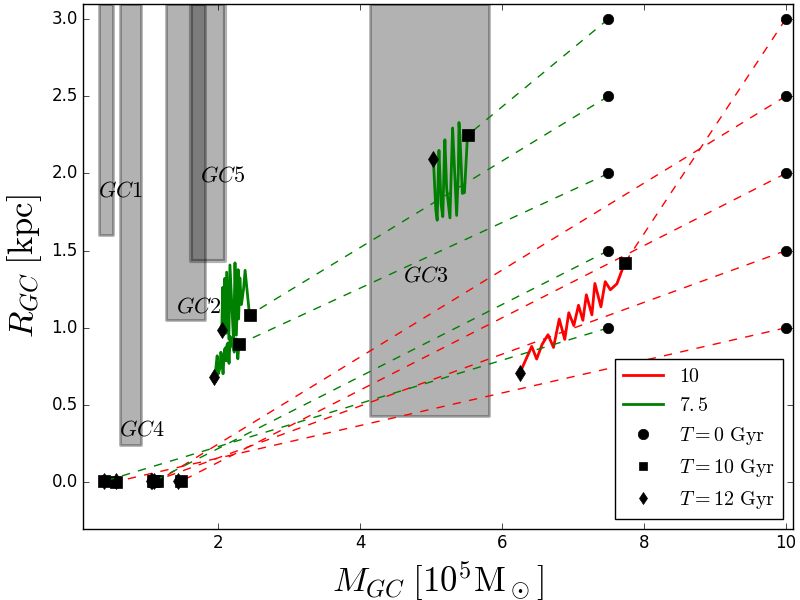}
\end{minipage} 
\begin{minipage}[t]{8.4cm}
\centering 
\includegraphics[width=8cm]{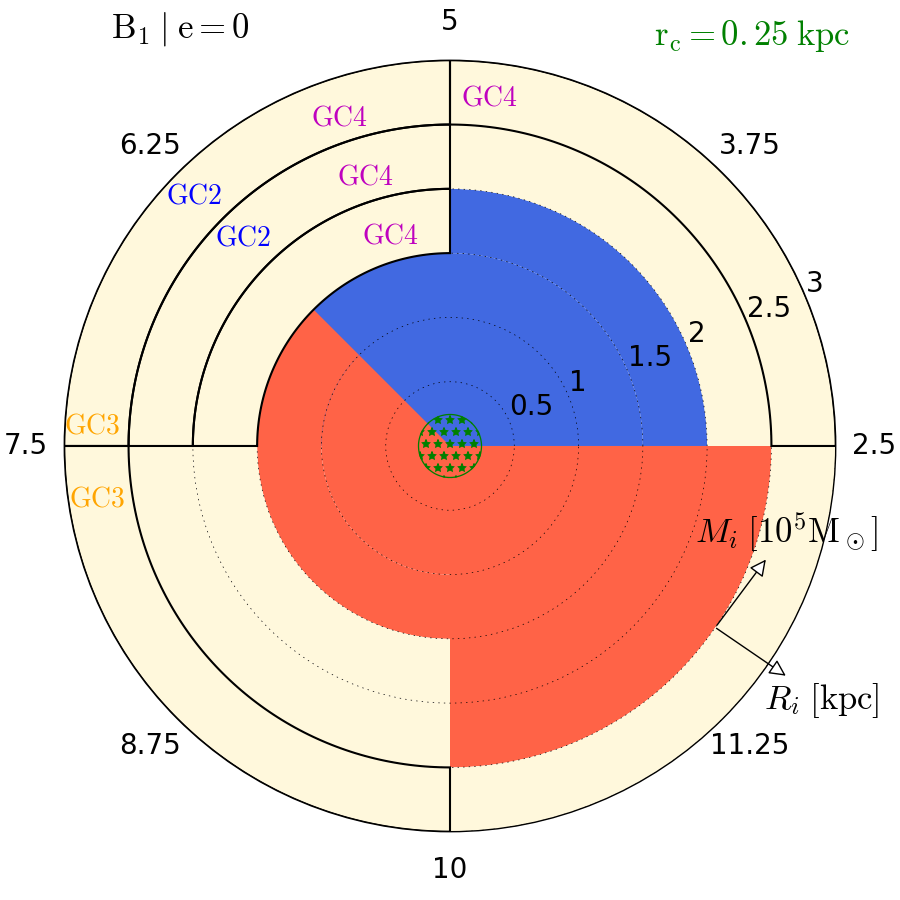}
\end{minipage}
\caption{Globular cluster orbital radius as a function of mass loss for B1 model with initial orbital radii $R_i$ = $[0.5, 1.0, 1.5, 2.0, 2.5]$ kpc, initial masses $M_i$ = $[2.5, 5.0, 7.5, 10]$ $\times10^{5}M_{\sun}$ and eccentricity parameter $e=0$. According to the uncertainty on globular cluster ages, we consider the positions and masses of the globular clusters between $T=$ 10 Gyr and $T=$ 12 Gyr. Cluster initial states ($T=0$ Gyr), represented by black points, are connected by dashed lines to final states ($T=10-12$ Gyr), represented by black squares and diamonds. Final states compatible with observations, i.e. projected distance and globular cluster mass with their uncertainties, are represented by a grey area for each globular cluster. The bottom left panel summarizes all results for B1 model. Blue (orange) regions represent the initial parameter range, where clusters were destroyed by the tidal field (have suffered from dynamical friction and have sunken to the galactic centre). The beige regions represent the range of initial parameters, where clusters can survive over 10-12 Gyr. Some of these regions are compatible with observed globular cluster distributions. We also represented the core region with green stars. Observed data for globular clusters are summarized in Table~\ref{go}. All halo parameters are summarized in Table~\ref{ta}.}
\label{gc}
\end{figure*}

We start the discussion of our results by presenting orbital radius as a function of mass loss for $B_1$ model for clusters with eccentricity parameter $e=0$, in greater detail. The first three plots of Figure~\ref{gc} depict the final radii and masses between 10 and 12 Gyr, due to the uncertainty on globular cluster ages, for clusters with initial orbital radii $R_i$ = $[0.5, 1.0, 1.5, 2.0, 2.5]$ kpc, initial masses $M_i$ = $[2.5, 5.0, 7.5, 10]$ $\times10^{5}M_{\sun}$. Final states compatible with cluster observed distributions, i.e. projected distance and mass with their uncertainties, are represented by a grey area for each globular cluster. Cluster initial states ($T=0$ Gyr), represented by black points, are connected by dashed lines to final states ($T=10-12$ Gyr), represented by black squares and diamonds. The right bottom panel of Figure~\ref{gc} summarizes all results for $B_1$ model with eccentricity parameter $e=0$ for the whole range of initial globular cluster orbital radii and masses considered in this work. Blue regions represent the initial parameter range, where clusters were destroyed by the tidal field. For instance, clusters with $2.5 \leq M_{i} \leq 5 \times10^{5}$M$_{\sun}$ started at $R_{i} \leq$ 2 kpc, will suffer from Fornax tidal field and be destroyed (see left upper panel of Figure~\ref{gc}). Only clusters, initialized at $R_{i} >$ 2 kpc, can survive. Thus, the beige regions represent the range of initial parameters, where clusters can survive over 10 - 12 Gyr. Some survival regions (beige regions) are compatible with observed globular cluster distributions. As an example, the left upper panel shows also that clusters with $5 \leq M_{i} \leq 7.5 \times10^{5}$M$_{\sun}$ and $2 \leq R_{i} \leq 3$ kpc can reproduce the observed mass and spatial distributions of GC2. Orange regions represent the initial parameter range, where clusters have suffered from dynamical friction and have sunken to the galactic centre. For instance, clusters with $7.5 \leq M_{i} \leq 10 \times10^{5}$M$_{\sun}$ started at $R_{i} \leq$ 1.5 kpc, will fall into Fornax centre (see left bottom panel). It is important to notice there is a transition from disrupted clusters to fallen clusters due to an initial mass increase for $5 \leq M_{i} \leq 7.5 \times10^{5}$M$_{\sun}$ and $R_{i} \leq 1.5$ kpc . We did not investigate sufficiently this initial range to know precisely the transition mass and radius. For $R_{i} < 1$ kpc, we expect the same final state, destroyed or fallen, than globular clusters started at $R_{i} = 1$ kpc, because dynamical friction and tidal effects are stronger at lower radii. The left bottom panel indicates also the dynamical behavior of clusters with $M_{i} \geq 10 \times10^{5}$M$_{\sun}$. All clusters with $M_{i} = 10 \times10^{5}$M$_{\sun}$ will sink to the centre of Fornax. According to these results, we expect that more massive clusters will also fall due to dynamical friction at these same initial radii. We also represented the core region with green stars (see right bottom panel of Figure~\ref{gc}). 

\begin{figure*}
\centering
\begin{minipage}[t]{8.4cm} 
\centering
\includegraphics[width=8cm]{pB1n.png}
\end{minipage} 
\begin{minipage}[t]{8.4cm}
\centering 
\includegraphics[width=8cm]{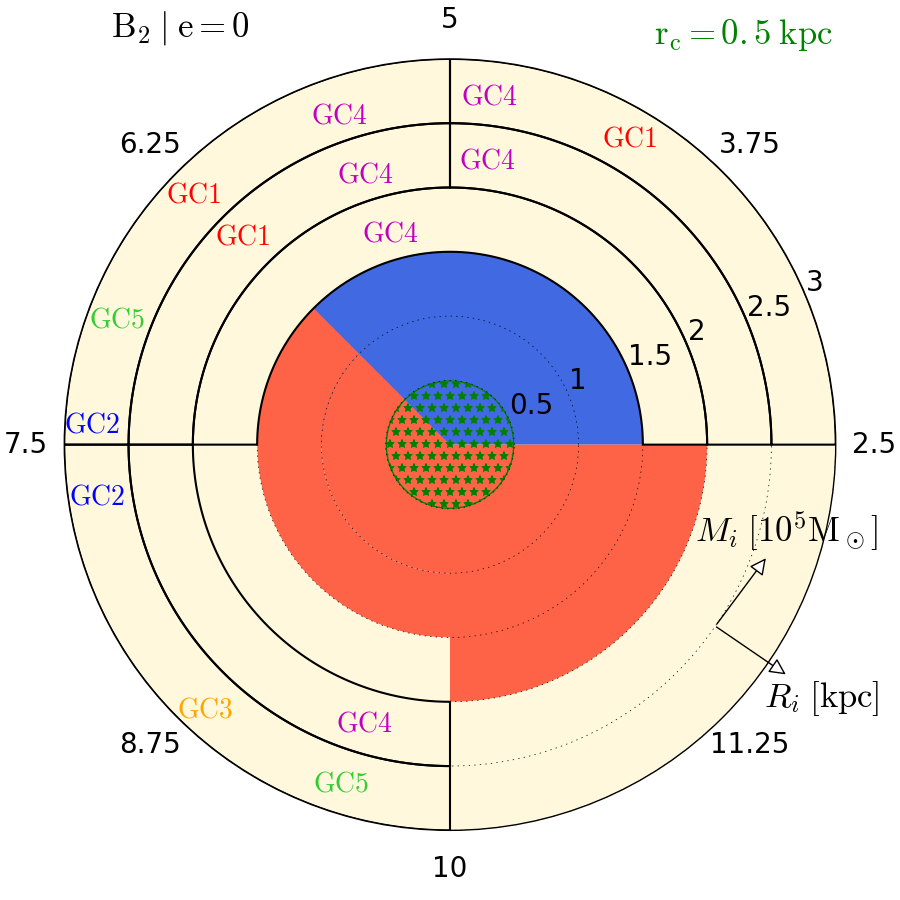}
\end{minipage}
\begin{minipage}[t]{8.4cm} 
\centering
\includegraphics[width=8cm]{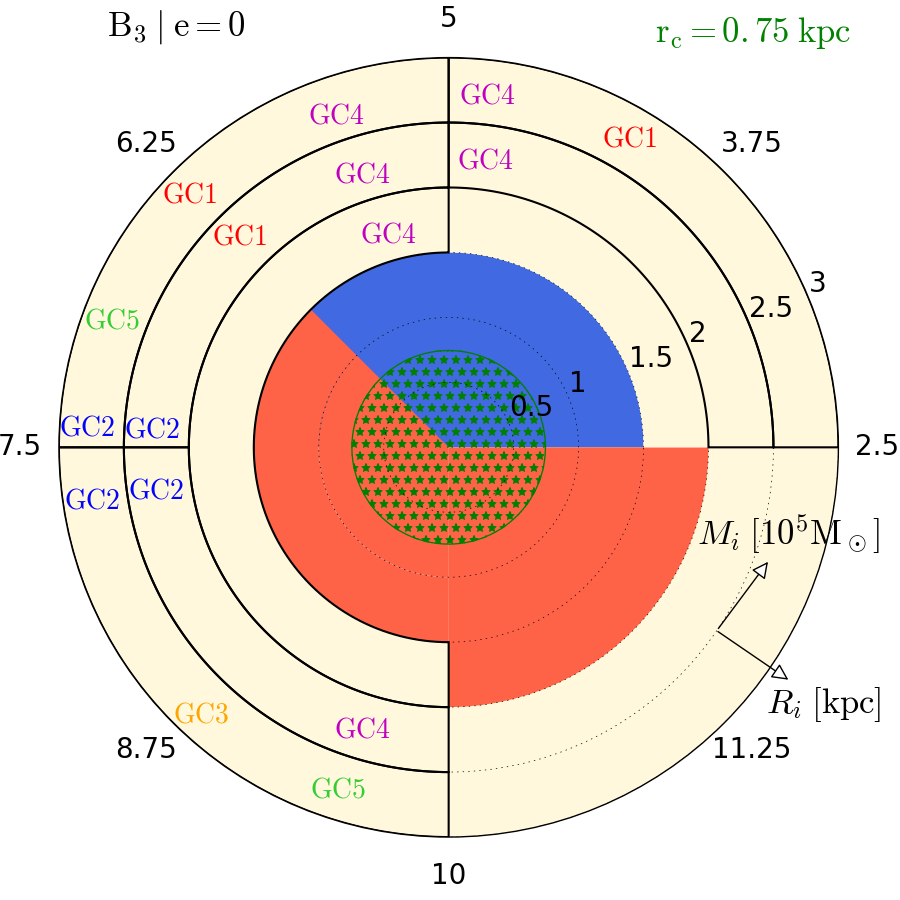}
\end{minipage} 
\begin{minipage}[t]{8.4cm}
\centering 
\includegraphics[width=8cm]{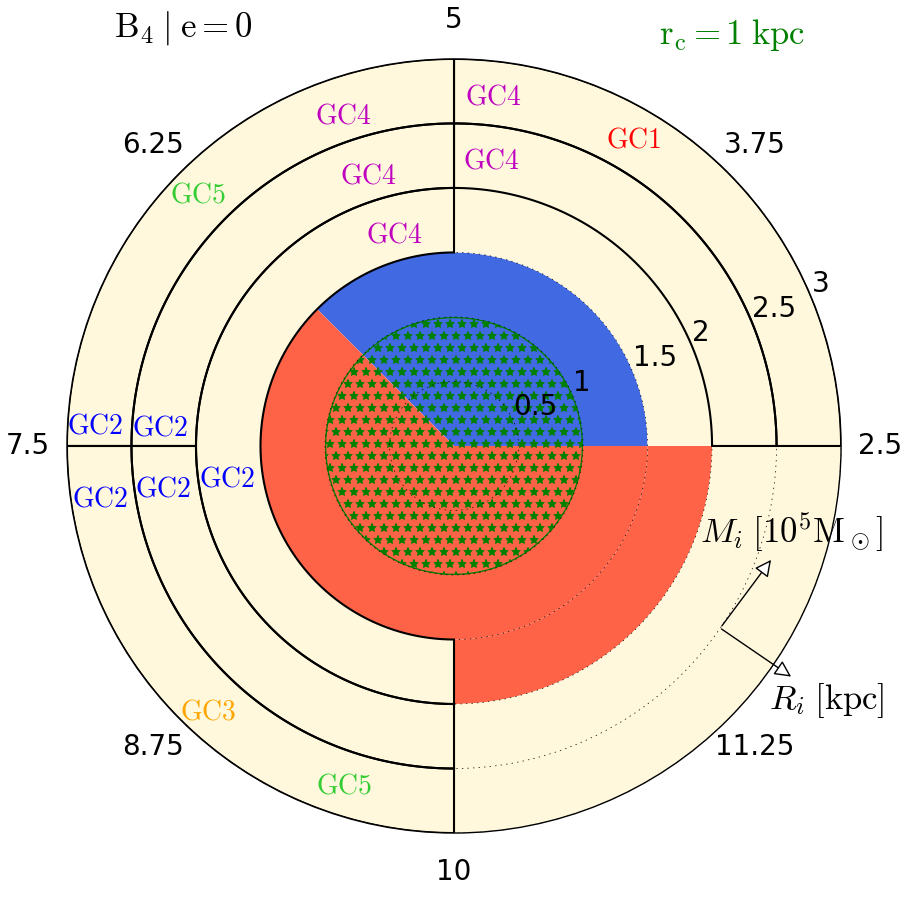}
\end{minipage}
\caption{Final states ($T=10-12$ Gyr) for initial globular cluster orbital radii $R_i$ = $[0.5, 1.0, 1.5, 2.0, 2.5]$ kpc and masses $M_i$ = $[2.5, 5.0, 7.5, 10]$ $\times10^{5}M_{\sun}$ with eccentricity parameter $e=0$ for cored halo models $B_n$. Blue (orange) regions represent the initial parameter range, where clusters were destroyed by the tidal field (have suffered from dynamical friction and have sunken to the galactic centre). The beige regions represent the range of initial parameters, where clusters can survive over 10-12 Gyr. Some of these regions are compatible with observed globular cluster distributions. We also represented the core region with green stars. Only B2, B3 and B4 models can reproduce the five globular clusters. Observed data for globular clusters and halo parameters are summarized in Table~\ref{go} and Table~\ref{ta}.}
\label{be0}
\end{figure*}

\begin{figure*}
\centering
\begin{minipage}[t]{8.4cm} 
\centering
\includegraphics[width=8cm]{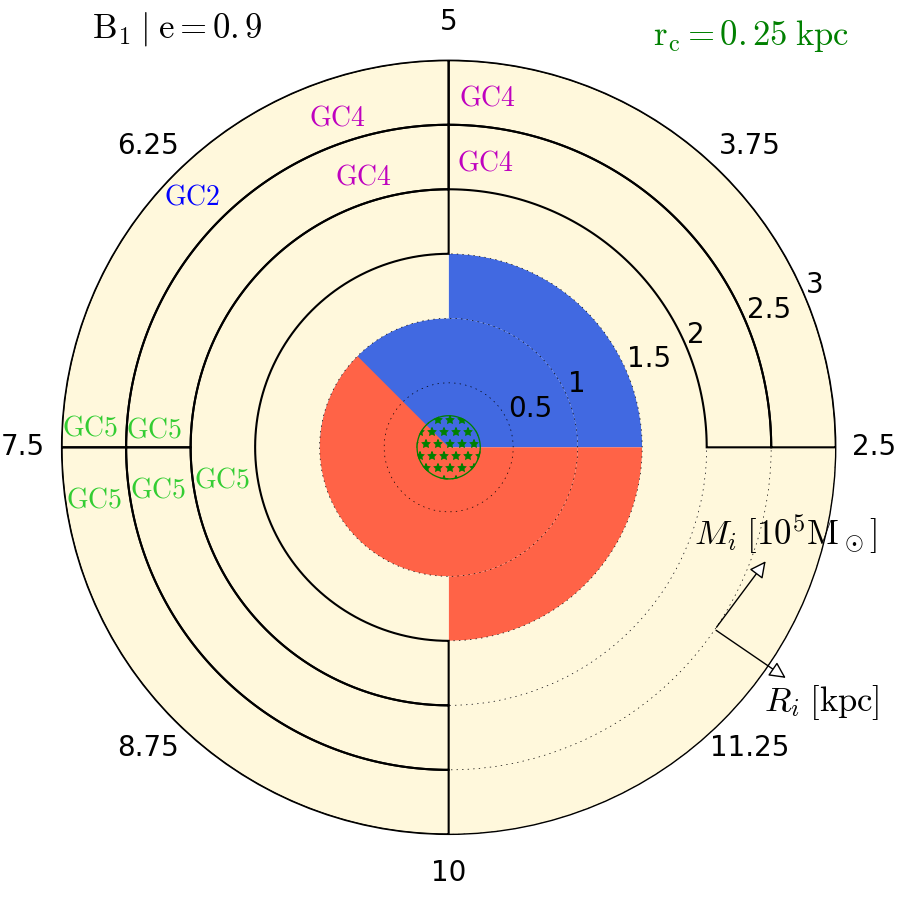}
\end{minipage} 
\begin{minipage}[t]{8.4cm}
\centering 
\includegraphics[width=8cm]{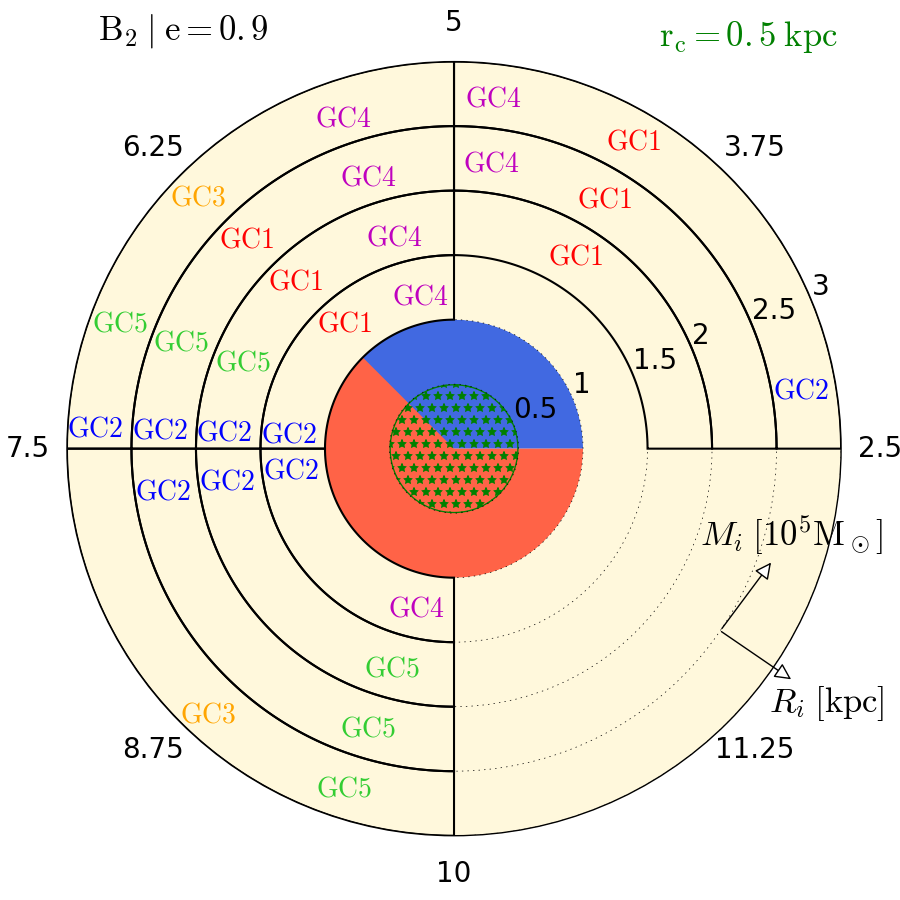}
\end{minipage}
\begin{minipage}[t]{8.4cm} 
\centering
\includegraphics[width=8cm]{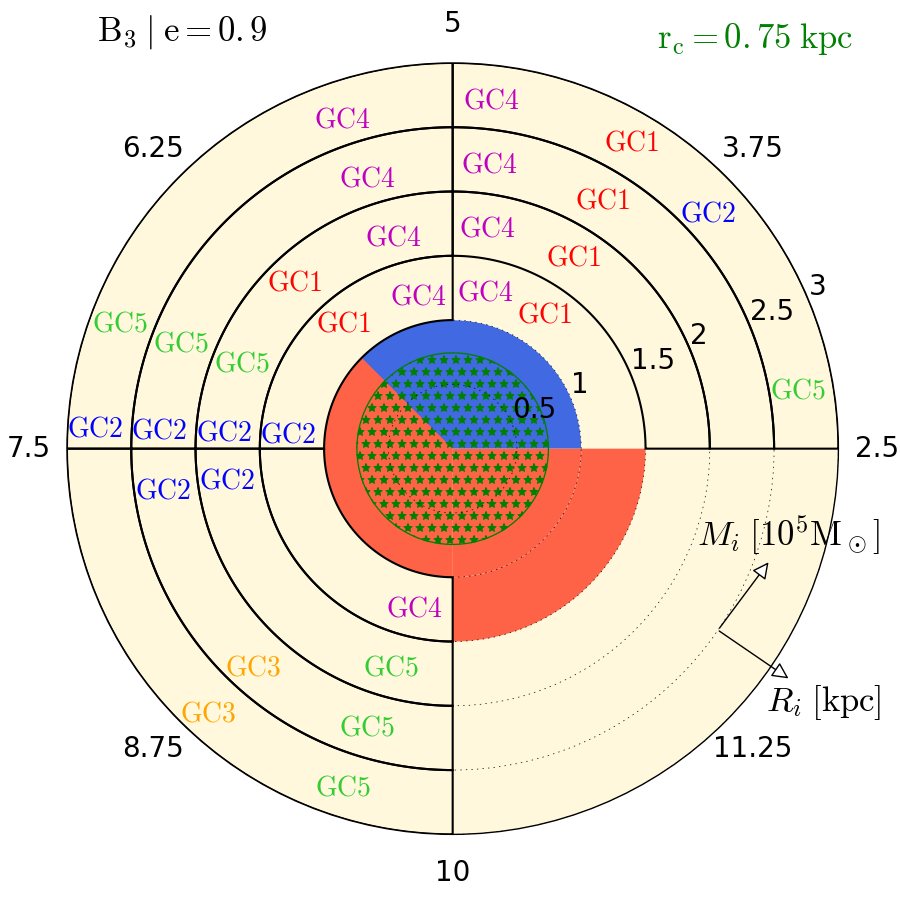}
\end{minipage} 
\begin{minipage}[t]{8.4cm}
\centering 
\includegraphics[width=8cm]{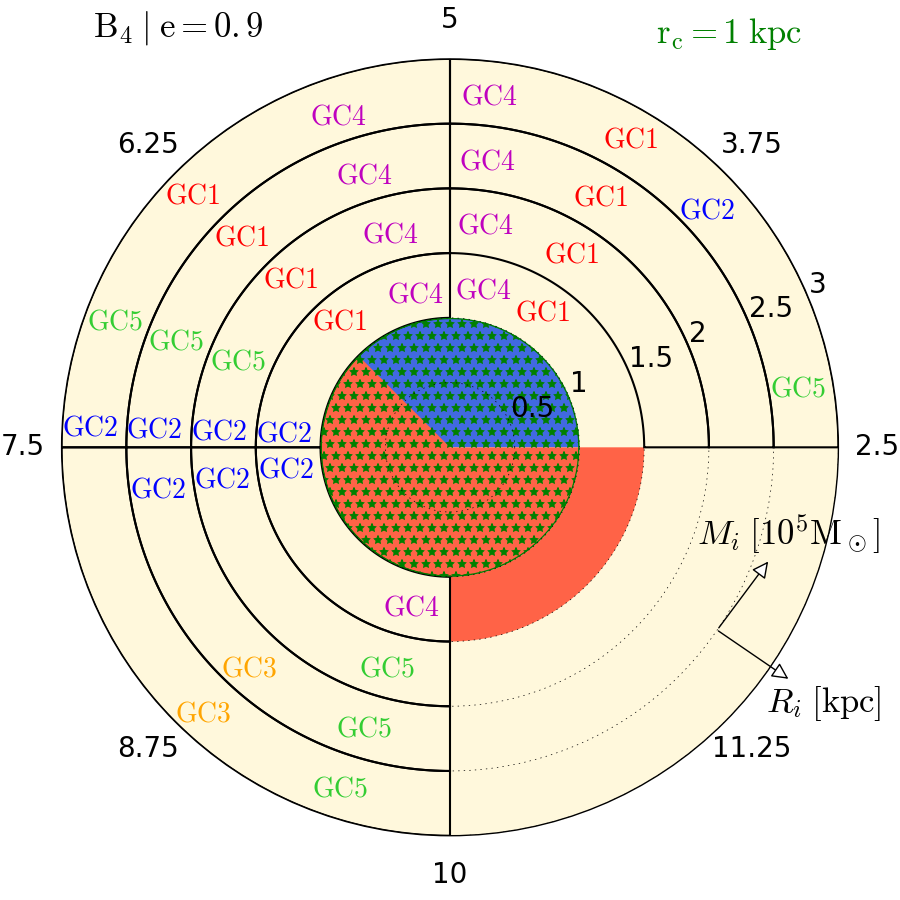}
\end{minipage}
\caption{Same as Figure~\ref{be0}, except with eccentricity parameter $e=0$ for cored halo models $B_n$. Only $B_2$, $B_3$ and $B_4$ models can reproduce the five globular clusters. Observed data for globular clusters and halo parameters are summarized in Table~\ref{go} and Table~\ref{ta}.}
\label{be9}
\end{figure*}

\begin{figure*}
\centering
\begin{minipage}[t]{8.4cm} 
\centering
\includegraphics[width=8cm]{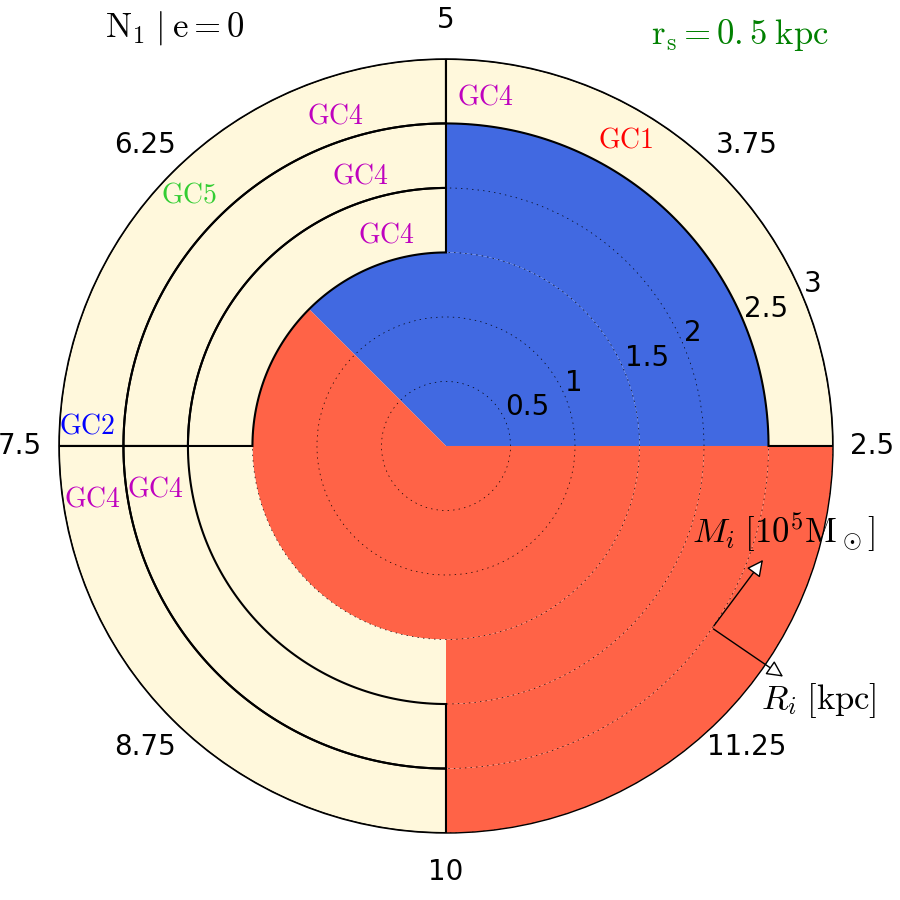}
\end{minipage} 
\begin{minipage}[t]{8.4cm}
\centering 
\includegraphics[width=8cm]{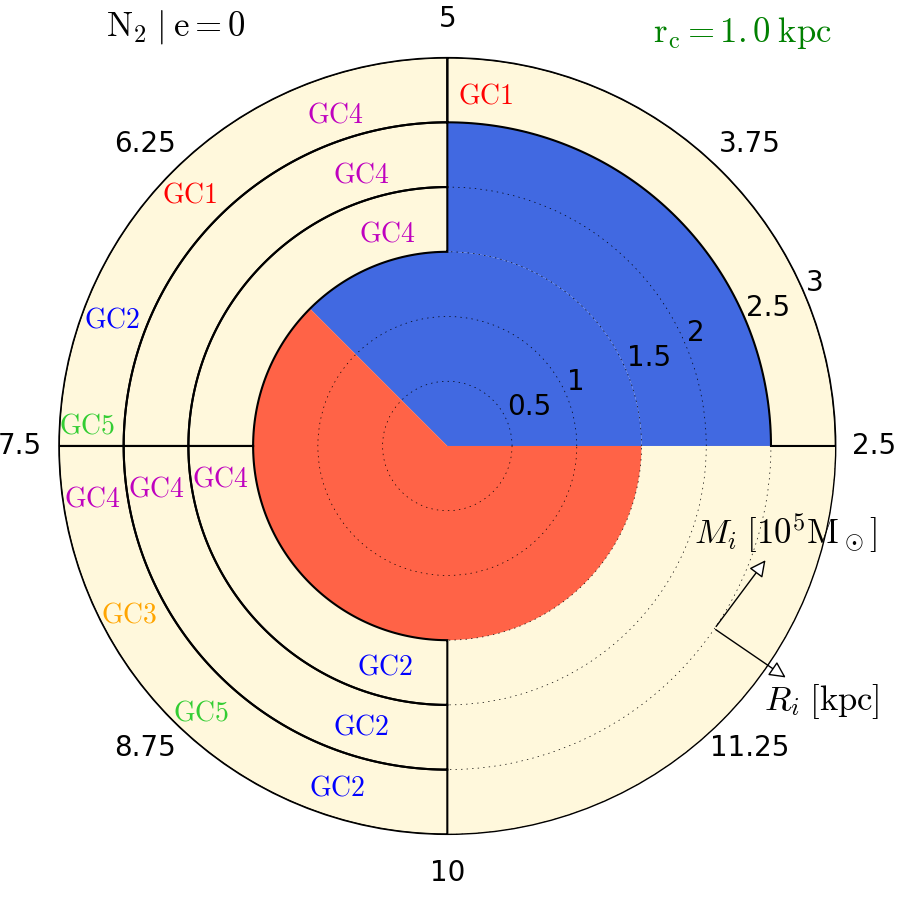}
\end{minipage}
\begin{minipage}[t]{8.4cm} 
\centering
\includegraphics[width=8cm]{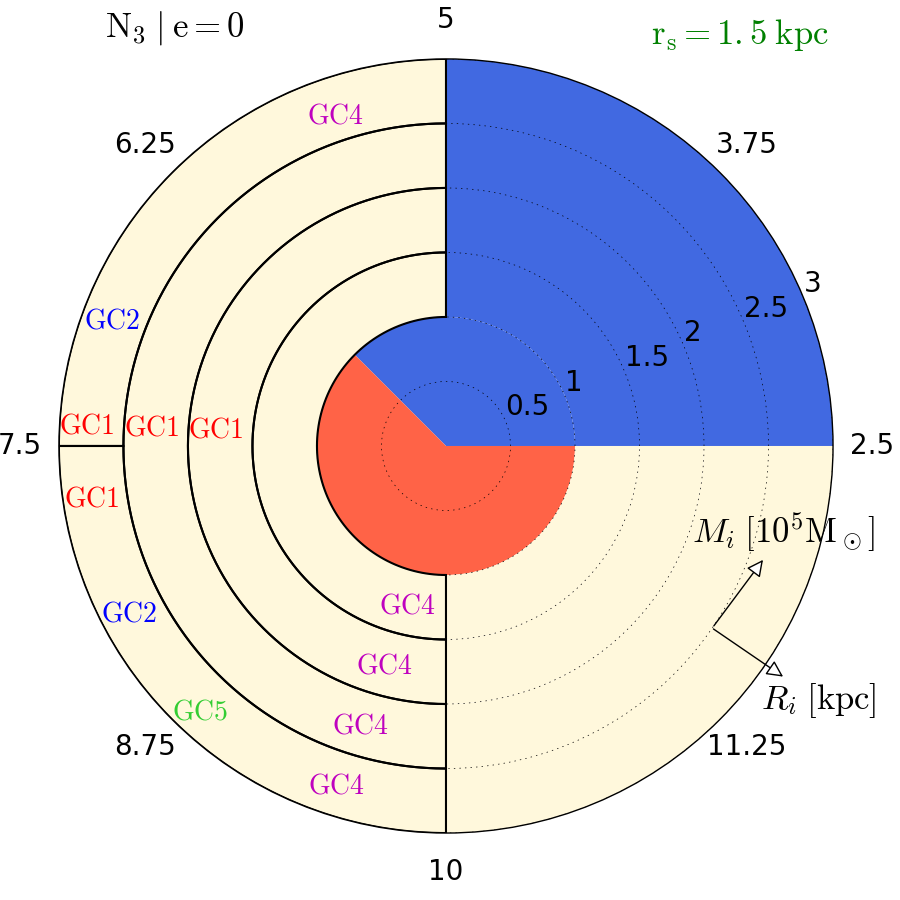}
\end{minipage} 
\begin{minipage}[t]{8.4cm}
\centering 
\includegraphics[width=8cm]{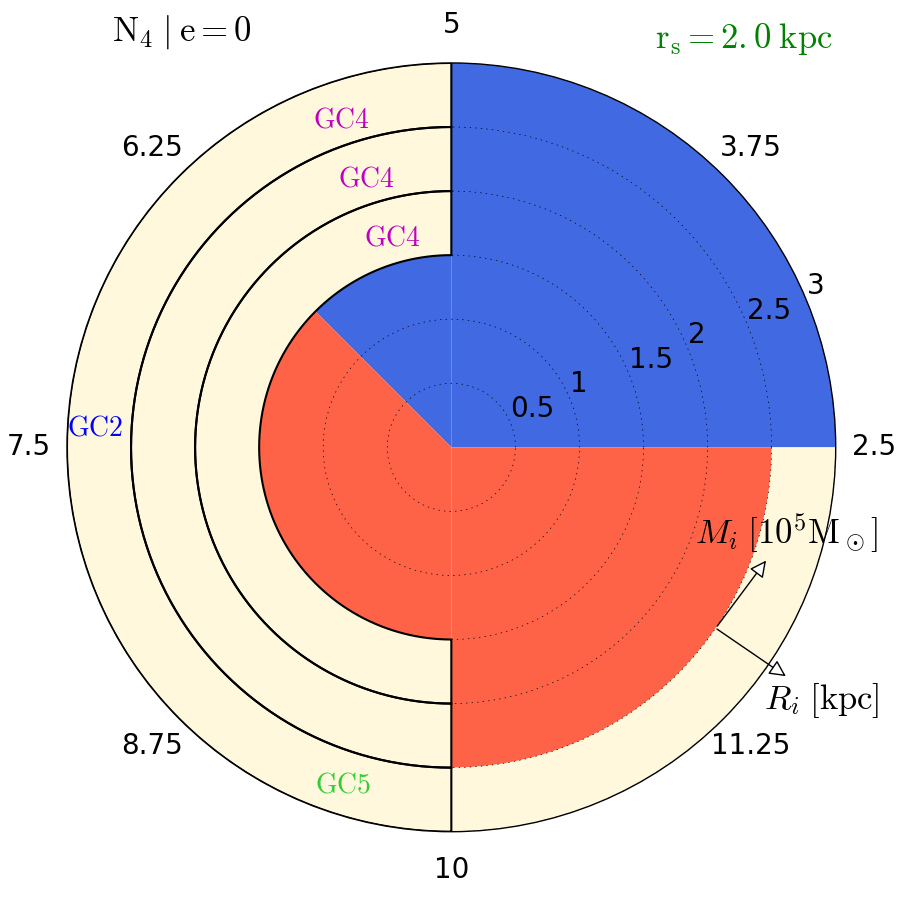}
\end{minipage}
\caption{Same as Figure~\ref{be0}, except with eccentricity parameter $e=0$ for cuspy halo models $N_n$. Only $N_2$ model can reproduce the five globular clusters. Observed data for globular clusters and halo parameters are summarized in Table~\ref{go} and Table~\ref{ta}.}
\label{Ne0}
\end{figure*}

\begin{figure*}
\centering
\begin{minipage}[t]{8.4cm} 
\centering
\includegraphics[width=8cm]{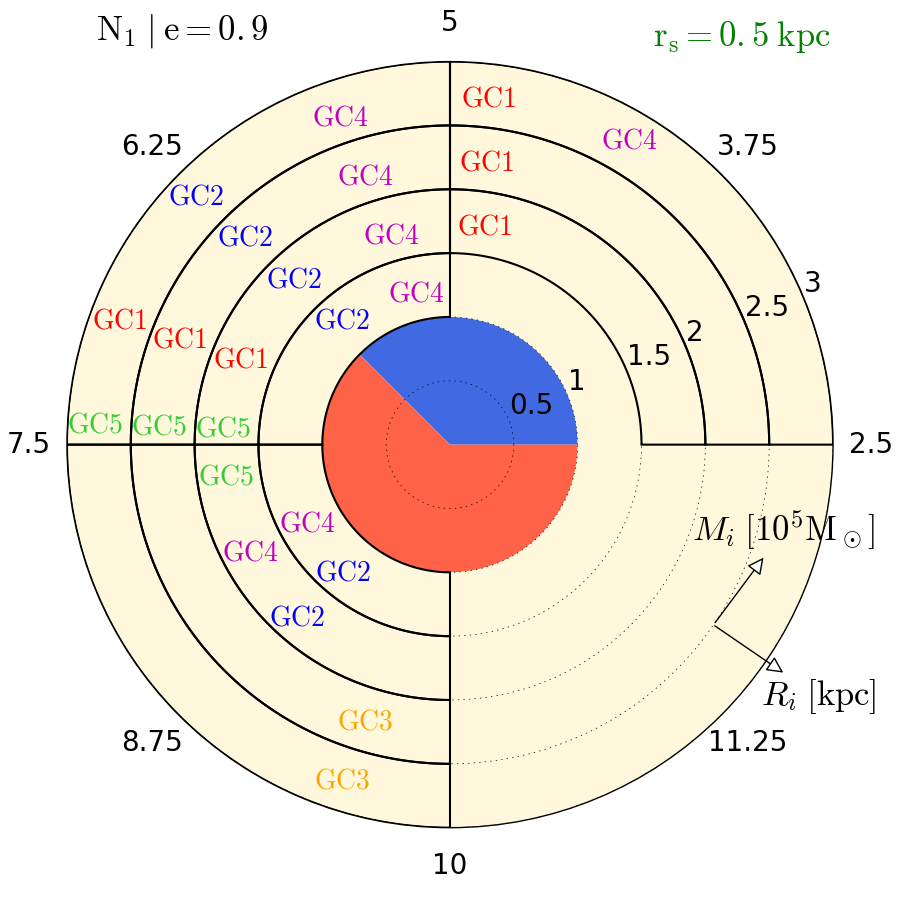}
\end{minipage} 
\begin{minipage}[t]{8.4cm}
\centering 
\includegraphics[width=8cm]{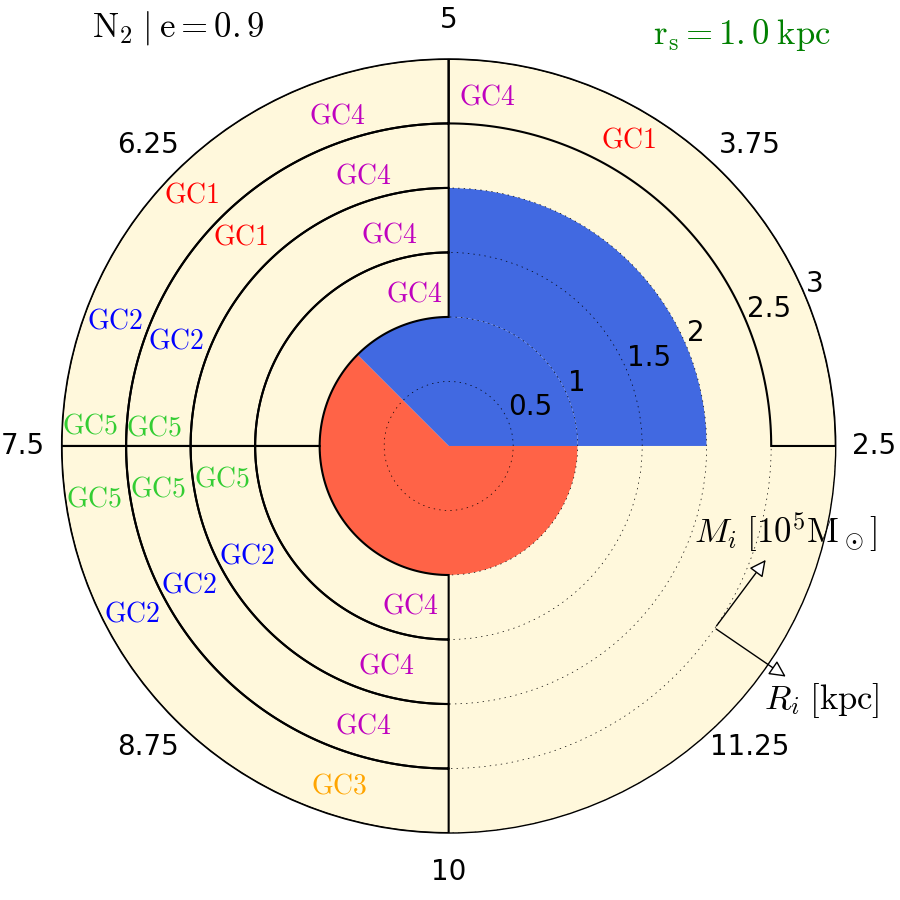}
\end{minipage}
\begin{minipage}[t]{8.4cm} 
\centering
\includegraphics[width=8cm]{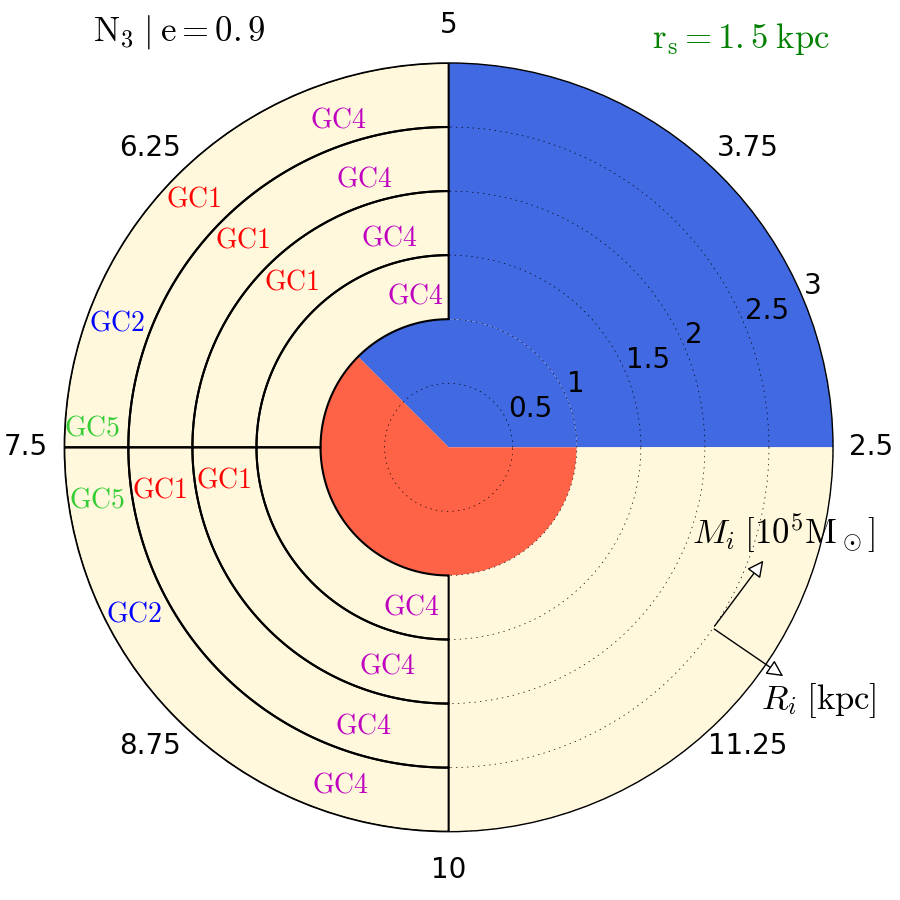}
\end{minipage} 
\begin{minipage}[t]{8.4cm}
\centering 
\includegraphics[width=8cm]{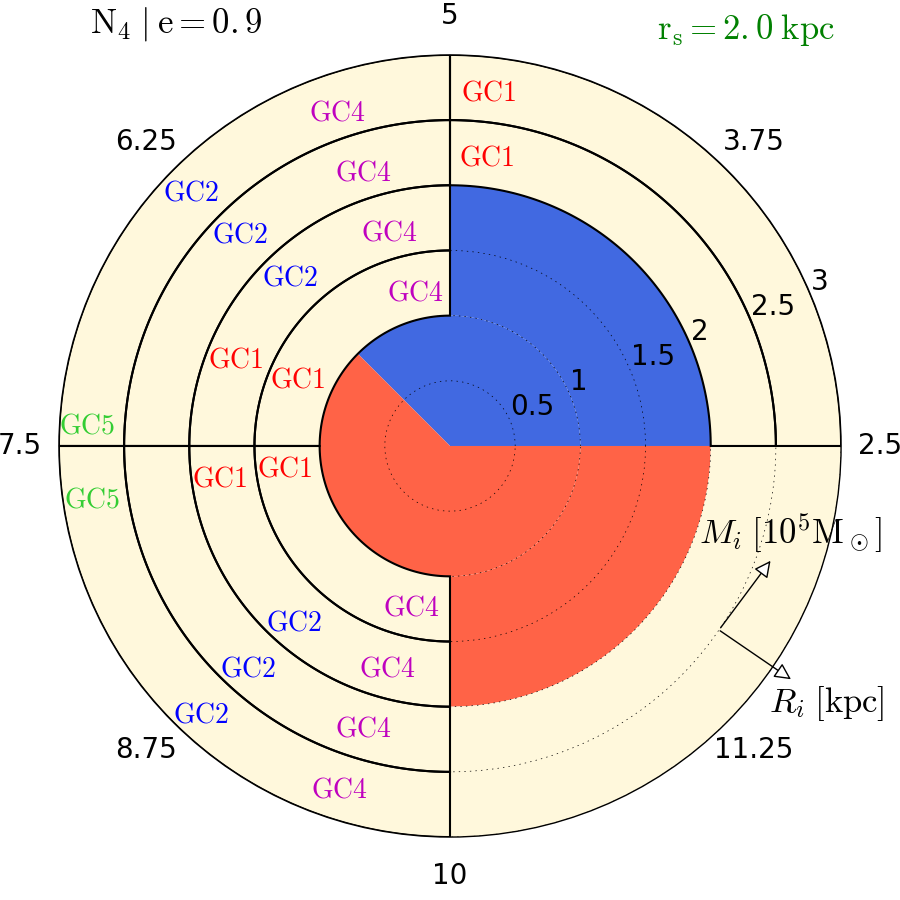}
\end{minipage}
\caption{Same as Figure~\ref{be0}, except with eccentricity parameter $e=0.9$ for cuspy halo models $N_n$. Only $N_1$ and $N_2$ models can reproduce the five globular clusters. Observed data for globular clusters and halo parameters are summarized in Table~\ref{go} and Table~\ref{ta}.}
\label{Ne9}
\end{figure*}

All our empirical results from our simulations can be summarized in Figure~\ref{sB} for cored halos, and Figure~\ref{sN} for cuspy halos. Hovewer, more details about globular cluster dynamic are presented in Figures~\ref{be0} and~\ref{be9} for cored halos, and Figures~\ref{Ne0} and~\ref{Ne9} for cuspy halos in order to highlight the range of initial parameters entailing that globular clusters will fall towards the centre of the galaxy (orange regions), be dissolved (blue regions) and survive within 10-12 Gyr (beige regions). We marked the globular clusters which are in agreement with survival regions. These figures compare the cored halo models $B_n$ and cuspy halo models $N_n$ for our range of initial cluster orbital radii and masses with an eccentricity parameter $e=0$ and $e=0.9$, respectively. 

Theoretically, we expect that low initial cluster mass entails that the globular clusters are destroyed by the tidal field . On the contrary, high initial cluster mass results in the globular cluster spiralling towards to the centre. Blue and orange regions represent clearly this expected dynamical behaviour for all halo models. Not surprisingly, we obtain more survival regions compatible with globular cluster observations for $e=0.9$ than for $e=0$, because clusters with high eccentric orbit are less affected by dynamical friction and tidal disruption. Indeed, these clusters spend too little time in high density regions. This result is valid for both cored and cuspy halos. For all halo models, we claimed that globular clusters have to be initially more massive in order to be in agreement with present observations. They need to be about 1.3 to 18 times more massive than the current clusters. We state also that initial orbital radii of clusters have to be at a distance greater than 1 kpc from the centre of Fornax for all halo models. In fact, clusters with $R_{i} < 1$ kpc are subject to a higher tidal disruption and dynamical friction, because they orbit in a high density region. 

$B_2$, $B_3$ and $B_4$ models hold for the five globular clusters of Fornax for both eccentric parameters with $B_1$ model being the sole exception. For $e=0$ ($e=0.9$), $B_1$ model is not valid for GC1 and GC5 (GC1 and GC3). GC1 proves to be the tightest constraint, ruling out $B_1$ model ($r_{c}=$ 0.25 kpc). In fact, this cluster requires at the same time a weak orbital decay and a very weak or huge mass loss induced by the dark matter halo. Concerning cuspy halos, simulations predict that $N_2$ ($r_{s}=$ 1 kpc) model can reproduce globular cluster distributions compatible with observations for both eccentric parameters (see Figure~\ref{sB}) . Otherwise, GC3 distributions remain a challenge for 
$N_1$, $N_3$ and $N_4$ models with our initial mass range. However, we expect that globular clusters with $M_{i} > 10\times10^{5}$M$_{\sun}$ could be compatible with GC3 observations for the $N_3$ ($r_{s}=$ 1.5 kpc) model (see Figures~\ref{Ne0} and~\ref{Ne9} for more details). Even if direct dynamical modelling of stellar population \citep{2011ApJ...742...20W,2013MNRAS.429L..89A} attest against the presence of a divergent cusp in Fornax, we find a cuspy halo, which can reproduce the observed distributions. Indeed, the NFW profile can be reconciled with observations.

Concerning cored halos, Figure~\ref{sB} imply that $B_2$, $B_3$ and $B_4$ models can reproduce observations for a relevant range of initial cluster orbital radii and masses. As observed above, $B_1$ model ($r_{c}=$ 0.25 kpc) cannot reproduce all the observed globular clusters, especially GC1 in both circular and high eccentric orbits (see Figures~\ref{be0} and~\ref{be9} for more details). In this context, using globular cluster distributions, it is possible to put constraints on the core radius. Thus, we derive a lower limit of $r_{c}\gtrsim$ 0.5 kpc. We ruled out the model with $r_{c}=$ 0.25 kpc, but we did not investigate core sizes between 0.5 and 0.25 kpc. We expect that the true lower limit is in this core size range. Our lower limit ($r_{c}\gtrsim$ 0.5 kpc) is in disagreement with \cite{2006ApJ...652..306S}, who found a upper limit of $r_{c}\lesssim$ 0.3 kpc, based on a constraint on central phase-space density of Fornax. However, \cite{2013MNRAS.429L..89A} showed that Fornax dark matter halo has a core with $r_{c}=1^{+0.8}_{-0.4}$ kpc by exploiting three distinct stellar subpopulations of Fornax. Their limit is totally compatible with our prediction from the $B_2$, $B_3$ and $B_4$ models (see Table~\ref{ta}). According to \cite{2011ApJ...742...20W}, the slope of the halo mass profile measured in Fornax suggests $r_{c}\gtrsim$ 1 kpc.

\section{Conclusions}

We have revisited the cusp-core problem applied to Fornax. Currently, this dSph galaxy has five globular clusters orbiting in its dark matter halo. Observational analyses suggest that the globular clusters were initially much more massive. For the first time, the Fornax globular system has been modelled with live objects, i.e.  self-gravitating systems only composed of star and dark matter particles, in order to properly implement dynamical friction and tidal effects between Fornax and globular clusters. We have performed 75 N-body simulations for cored and cuspy halos in Milky Way tidal field modelled by a static potential. 

Using constraints from globular cluster spatial and mass distributions, we showed that Fornax can have either a cored or a cuspy halo. More precisely, our results have revealed a lower limit of a core size of $r_{c}\gtrsim$ 500 pc for Fornax cored halo. Even if many studies attest against the presence of a divergent cusp in Fornax, we show also that, from our globular cluster constraints, Fornax can have a cuspy halo described by $N_2$ model. 

Apart from CDM, all variant dark matter theories, including warm dark matter (WDM), fuzzy dark matter (FDM) and self-interacting dark matter (SIDM) are apparently in favour of cored halos. If the halo of Fornax is composed of WDM, \cite{2006ApJ...652..306S} established that $r_{c}\lesssim$ 85 pc in order to avoid conservative limits from the Ly$\alpha$ forest power spectrum. From FDM simulations, \cite{2018ApJ...853...51Z} showed in general that a solitonic core with a size of 3 kpc emerges, composed of dark matter particles, which are ultra-light axions with a mass $\mathcal{O}$($10^{-22}$) eV. From SIDM simulations, \cite{2013MNRAS.431L..20Z} imposed that the Fornax core radius should be $r_{c}>$ 500 pc. This estimate is based on the circular velocity of Fornax. Future work could investigate the dynamical behaviour of globular clusters in these theories, in particular the impact on dynamical friction.

\section{Acknowledgments}
We thank our anonymous referee for the helpful comments and suggestions that improved our work. We thank Raphael Sadoun, Miki Yohei, Alexander Wagner, Guilhem Lavaux, Yves Revaz, Volker Springel, Pierre Salati, Simon White, Gary Mamon, Stephane Charlot and Stephane Colombi for their helpful suggestions concerning simulations and stimulating conversations. We would also like to thank David Valls-Gabaud, Dante Von Einzbern, Apolline Guillot and my cats (Totem $\&$ Tabou) for their constructive suggestions to improve the manuscript. This work was supported by the European Union's Horizon 2020 research and innovation program under the Marie Sklodowska-Curie grant agreement No 690904 and it has made use of the Horizon Cluster hosted at the Institut d'Astrophysique de Paris.

\appendix

\section{Numerical convergence}

\begin{figure}
\centering
\includegraphics[width=0.47\textwidth]{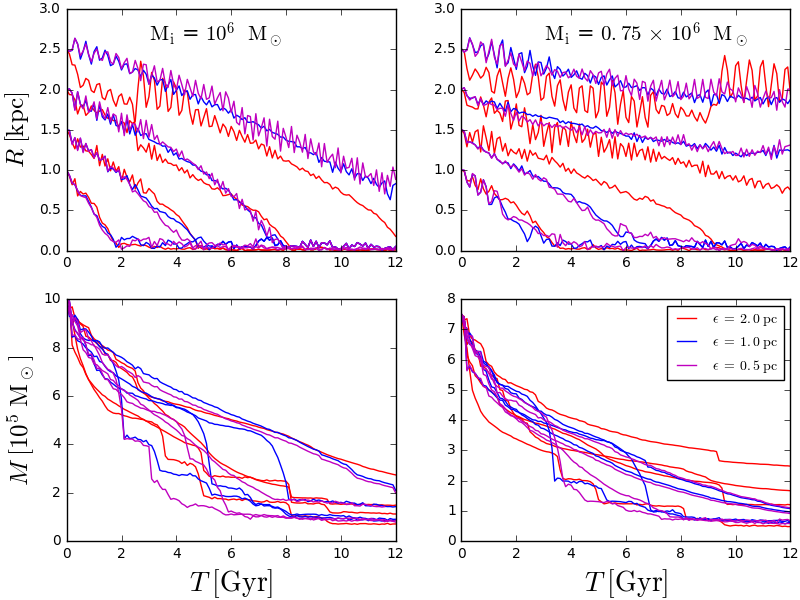}
\caption{Impact of softening length on the radial and tidal evolutions of globular clusters on a circular orbit with initial mass $M_i$ = 0.75 and 1 $\times10^{6}M_{\sun}$ for $B_4$ model. Top and bottom panels show the evolution of orbital radius and cluster mass, respectively. We ran simulations with three different softening lengths $\epsilon$ = [0.5, 1, 2] pc in order to ensure that our simulations do not suffer from numerical noise. Our simulations are well converged between 10 and 12 Gyr for $\epsilon$ = 0.5 and 1 pc. We chose $\epsilon$ = 1 pc as the softening length for all our simulations.}
\label{soft}
\end{figure}

\begin{figure}
\centering
\includegraphics[width=0.47\textwidth]{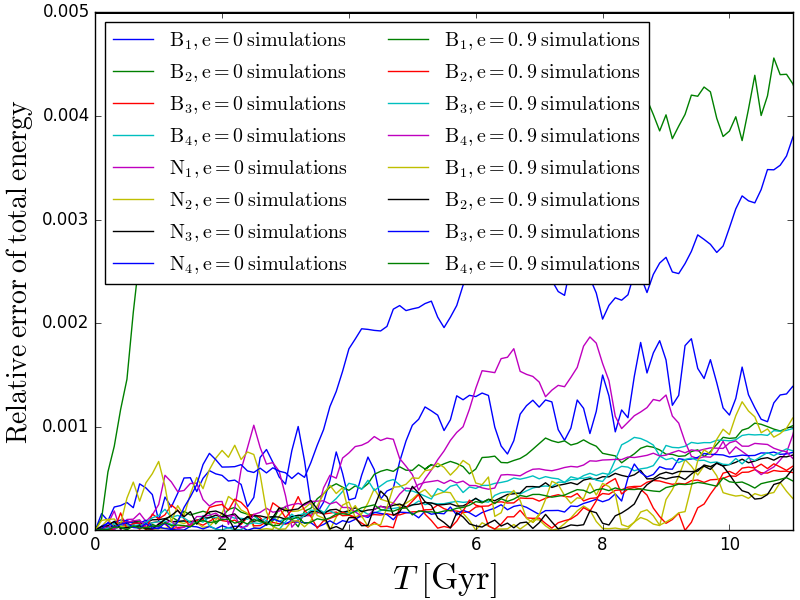}
\caption{Relative error of the total energy of the system over the time for all halos models for both eccentric parameters. The energy of the system is conserved in all our simulations because the energy relative errors are lower than 5$\%$.}
\label{ener}
\end{figure}

In this section, we assess the impact of a numerical parameter that controls the accuracy of our simulations; the softening length $\epsilon_0$ = 1 pc. To test how the softening length impacts on the radial and tidal evolution of globular clusters, we ran simulations with three different softening lengths $\epsilon$ = $\epsilon_0$/2, $\epsilon_0$ and 2$\epsilon_0$ in order to ensure that our simulations do not suffer from numerical noise. The evolution of the orbital radius and the mass loss of five clusters moving on a circular orbit in model $B_4$ (see Table~\ref{ta}) is shown in Fig~\ref{soft} for three different softening lengths. It can be seen that orbital orbital and tidal evolutions are very similar for $\epsilon$ = 0.5 and 1 pc. However, for $\epsilon$ = 2 pc, numerical noise causes artificial decay and enhanced disruption of the cluster. Our simulations are well converged between 10 and 12 Gyr for $\epsilon$ = 0.5 and 1 pc. We chose $\epsilon$ = 1 pc as the softening length for all our simulations. From our simulations, we can provide the relative error of total energy of the system in Fig~\ref{ener}. For all runs, the energy relative error is lower than 5$\%$. We state that the energy of the system is conserved in all our simulations.

\section{Milky Way tidal field}

\begin{figure}
\centering
\includegraphics[width=0.47\textwidth]{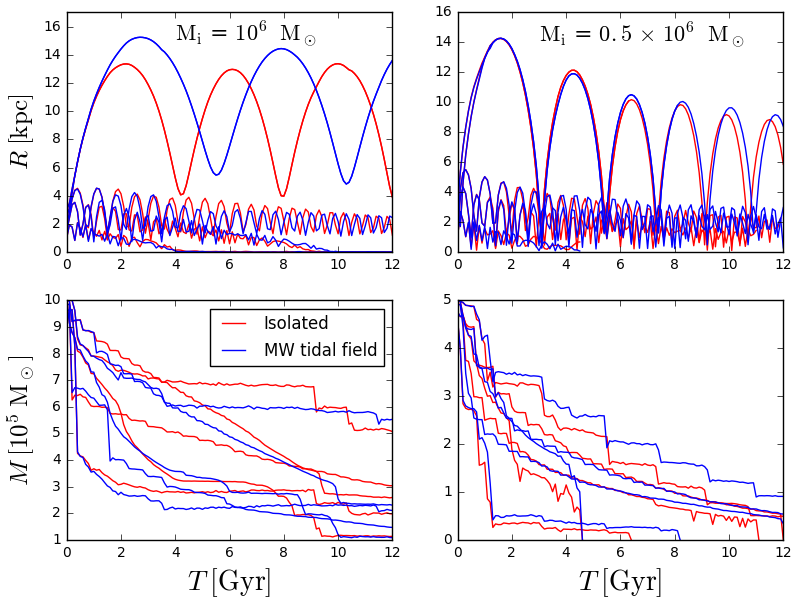}
\caption{Impact of MW tidal field on the radial and tidal evolution of globular clusters on a eccentric orbit ($e=0.9$) with initial mass $M_i$ = 0.5 and 1 $\times10^{6}M_{\sun}$ for $B_1$ model. Top and bottom panels show the evolution of orbital radius and cluster mass, respectively. We ran simulations with and without static potential in order to measure the importance of MW tidal field. MW tidal field was taken into account in our simulations, especially for high eccentric orbit}
\label{mw}
\end{figure}

Our host galaxy potential is modelled from \cite{1991RMxAA..22..255A} consists of a stellar bulge as a Plummer sphere \citep{1911MNRAS..71..460P}, a disc represented by the potential from \citep{1975PASJ...27..533M} and a spherical dark matter halo described by NFW profile \citep{1997ApJ...490..493N}. For this model, we used the revised parameters from \cite{2013A&A...549A.137I} (see their Table 1). Nevertheless, MW tidal field can have a significant impact on globular clusters on eccentric orbits. In Figure~\ref{mw}, we represented the orbital radius and mass of globular clusters on an eccentric orbit ($e=0.9$) with initial mass $M_i$ = 0.5 and 1 $\times10^{6}M_{\sun}$as a function of time for $B_1$ model. It can be seen in this figure that the orbits tend to be spread by the MW tidal field. In other words, the apocentre and pericentre are increased by the Galactic field. This effect becomes more important as the ratio between the initial orbital radius and the tidal radius increases. Thus, MW tidal field was taken into account in our simulations, especially for high eccentric orbit.

\bsp	
\label{lastpage}
\end{document}